\documentstyle[12pt]{article}
\title{\hfill \fbox{\small UnB.FIS.FM-002/92}\\
\vspace{2cm}
Topological Defects and Corrections to the Nambu Action}
\author{Vanda Silveira\thanks{Email  Vanda@BRUNB} \\Departamento de
Fisica\\Universidade de Brasilia\\70910 Brasilia,D.F.Brazil\\
\and
M.D.Maia\thanks{Email Maia@BRUNB} \\Departamento de Matematica\\
Universidade de Brasilia\\70910 Brasilia,D.F. Brazil}

\begin{document}

\begin{abstract}
The effective action of a (1+2)-dimensional defect is obtained as an
expansion in powers of the thickness.Considering non-straight solutions
as the zero order term, the corrections to the Nambu action are found to 
depend on the curvature scalar and on the gaussian curvature .   
\end {abstract}

\newpage

\section{Introduction}
Topological defects can arise in field theories with a non-trivial vacuum
manifold. The dynamical evolution of these objects can be derived from the
field theory,  following the time evolution of the fields. However, in the
absence of explicit solutions, effective actions can be constructed for 
strongly locallized defects, describing directly the geometrical evolution 
of the defect world-surface.

Considering the straight and static solution as the zero order limit in the
thickness $\epsilon$, and expanding the equations in powers of $\epsilon$,
it was proved that defects have generalized Nambu actions [1] in the lower 
order of the characteristic width of the defect. Higher order terms were
expected to include corrections to the Nambu action [2].

However, R.Gregory  showed, recently, that if the zero order solutions 
are consistently taken into account in the second order corrections, then
these
corrections to the Nambu action are proportional to R, the Ricci curvature
[3,4].
This term vanishes for particles and it is a topological constant for strings.

Only bubbles and higher-dimensional defects would have a non trivial  
contribution to the dynamical equation.

One important ingredient in R.Gregory's analysis is the starting point: the
static and straight solution as the solution to represent the limit 
$\epsilon \rightarrow 0$. This means that two limits are taken simultaneously
$\epsilon \rightarrow 0$ and $\rho \rightarrow \infty$, in which $\rho$ stands
for an average or local curvature of the defect surface.

Nevertheless, there is no reason to reject a treatment in which 
$\epsilon \rightarrow 0$, or better $\frac{\epsilon}{\rho} \rightarrow 0$ as
the zero order limit, without requiring the $\rho \rightarrow \infty $ limit.
This
alternative procedure would be more suitable to describe, for example, the  
evolution of circular strings and spherical bubbles, that may be thin, but are
neither static nor straight. 

Roughly, the importance of the curved features of the 
solutions will depend on the relative sizes of terms like $\phi^{''}$ and
$\frac{1}{\sqrt{-g}}(\sqrt{-g})^{'} \phi^{'}$, in which $\phi$ is the scalar
field and ' stands for derivative in a direction perpendicular to the defect 
surface. This contribution may be important or not, but it is independent
of the defect width.

In what follows, we consider the simplest case of a scalar field with two
vacuum states. We start with a non-straight solution with a domain wall, and, 
in the zero order limit, we recover the Nambu action. The next order 
correction is also calculated and Gregory's term is obtained together with
extra terms. 

\section{Gaussian Coordinates}
 
We consider a scalar field $\phi$ and a potential $V(\phi)$ with symmetry
breakdown and two ( or more) vacuum states $\phi_{1}$ and $\phi_{2}$. We
select solutions with a topological defect, and they are considered highly
concentrated on a well defined surface. If V($\phi$) is the usual
$\lambda (\phi^{2} - v^{2})^{2}$ potential than this surface may be identified
with $\phi = 0$ . 
 
A new coordinate system based on the defect surface is introduced. 
In the general case, a p-dimensional defect spans a (p+1)-dimensional 
manifold in the space-time.This world surface is locallyzed by
$X^{\mu}(\sigma^{A})$ with $\mu = 0,1...n$, and $\sigma^{A}$ are the
coordinates on the surface with $A= 0,...p$. 
$X^{\mu}_{,A} = \frac{\partial X^{\mu}}{\partial \sigma^{A}}$ are tangent
vectors. 
On each point of the surface, there is a normal plane, spanned by the normal
vectors $N^{\mu}_{i}$ with $i = 1,...m$ and $p + m = n$.  In each normal
direction, we choose coordinates $\xi^{i}$. Normal vectors are
normalized and orthogonal to the tangent vectors. 
Near the surface, the points are
locallized by:

\begin{equation}
Z^{\mu} (\sigma^{A},\xi^{i}) = X^{\mu}(\sigma^{A}) +
\xi^{i} N^{\mu}_{i}(\sigma^{A})             
\end{equation}

We start with $G_{\mu\nu}$, a diagonal metric of the flat (n+1)-dimensional
space that may be Minkowski metric. In the new coordinate system, we have
$g_{\mu\nu}$ with :

\begin{equation}
G_{\mu\nu} dZ^{\mu} dZ^{\nu} = g_{A B} d\sigma^{A} d\sigma^{B} +
                               2 g_{A j} d\sigma^{A} d\xi^{j} +
							   g_{i j} d\xi^{i} d\xi^{j}
\end{equation}

{}From (1), we have $dZ^{\mu}$ given by :

\begin{equation}
dZ^{\mu} = (X^{\mu}_{,A} + \xi^{i} N^{\mu}_{i,A})d\sigma^{A} +
           N^{\mu}_{i} d\xi^{i}
\end{equation}

and replacing (3) in the left-hand side of (2), we get $g_{\mu\nu}$ :

\begin{equation}
g_{AB} = \tilde{g}_{AB} + \xi^{i} N^{\mu}_{i,A} X_{\mu,B} +
         \xi^{j} N^{\nu}_{j,B} X_{\nu,A} +
		 \xi^{i} \xi^{j} N^{\mu}_{i,A} N_{j \mu,B}
\end{equation}

\begin{equation}
g_{iB} = \xi^{j} N^{\mu}_{i} N_{\mu j,B}
\end{equation}

\begin{equation}
g_{Aj} = \xi^{i} N^{\mu}_{i,A} N_{\mu j}
\end{equation}
		 
\begin{equation}
g_{ij} = \delta_{ij}
\end{equation}

for $\tilde{g}_{AB} = G_{\mu\nu} X^{\mu}_{,A} X^{\nu}_{,B}$ and 
$N^{\mu}_{i} N_{\mu j} = \delta_{ij}$. The inverse metric $g^{\mu\nu}$
can be computed in powers of $\xi^{i}$.Requiring that
$g_{\mu\nu} g^{\nu\lambda} = \delta_{\mu}^{\lambda}$ , we find :

\begin{equation}
g^{AB} = \tilde{g}^{AB} - \xi^{i} N^{\mu,A}_{i} X^{,B}_{\mu} -
                  \xi^{i} N^{\mu,B}_{i} X^{,A}_{\mu} + O(\xi^{2})
\end{equation}

\begin{equation}
g^{iB} = - \xi^{j} N^{\mu i} N_{\mu j}^{,B} + O(\xi^{2})
\end{equation}

\begin{equation}
g^{Aj} = - \xi^{i} N^{\mu ,A}_{i} N^{j}_{\mu} + O(\xi^{2})
\end{equation}

\begin{equation}
g^{ij} = \delta^{ij} + 
\xi^{k} \xi^{l} N_{\mu k ,A} N^{\mu i} N_{\nu l}^{ ,A} N^{\nu j}  +
  O(\xi^{3})
\end{equation}  				  

The crossed terms $g_{iB}$ and $g_{Aj}$  are related to 
$A_{ijB} = N_{i}^{\mu} N_{\mu j,B}$ , which is the torsion or twisting
vector and vanishes in the case of hypersurfaces (n= p+1) [5] ,
 making :

\[ g_{iB} = g_{Aj} = 0 \]

\begin{equation}
  g^{iB} = O(\xi^{2}),  g^{Aj} = O(\xi^{2}) , g^{ij} = O(\xi^{3})
\end{equation}  

\section{Solutions in powers of the defect width}

The action of a field configuration is given by :

\begin{equation}
S= \int \sqrt{-G}{\cal L} \, d^{n+1}Z = 
\int \sqrt{-g} {\cal L}\, d^{p+1}\sigma \, d^{m}\xi
\end{equation}

with ${\cal L} = \frac{1}{2} \partial^{\mu}\phi \partial_{\mu}\phi - V(\phi)$.

To construct
the effective action, we consider solutions for $\phi$ with a topological
defect and replace it in (13). This new action is, then, understood as
function
of the surface coordinates, and the dynamical evolution is studied directly
for $X^{\mu}(\sigma^{A})$.

The defect surface is characterized by a constant value of $\phi$ , namely
$\phi = 0 $ and the coordinates are chosen with $\xi^{i} = 0$ on this surface.
The  equation of motion from (13) is then written as :

\[
\frac{1}{\sqrt{-g}} \partial_{i}(\sqrt{-g} g^{ij} \partial_{j} \phi) +
\frac{1}{\sqrt{-g}} \partial_{i}(\sqrt{-g} g^{iA} \partial_{A} \phi) + 
\frac{1}{\sqrt{-g}} \partial_{A}(\sqrt{-g} g^{Aj} \partial_{j} \phi) +
\]

\begin{equation}
\ + \frac{1}{\sqrt{-g}} \partial_{A} (\sqrt{-g} g^{AB} \partial_{B}
\phi) + 
\frac{\partial V}{\partial \phi} = 0
\end{equation}

Since the solutions we are interested on are concentrated around $\xi^{i} = 0$,

both $\phi - v$ and $\partial_{\mu} \phi$ decrease fast with $\xi^{i}$.
Because of this, it is reasonable to solve (14) in powers of 
the typical thickness of the defect core ($\epsilon$). 
We consider solutions of the form :
 
\[  \phi = \phi_{0}(\xi^{i}) +  \phi_{1}(\xi^{i}) +  \phi_{2}  \] 

with $\phi_{1}$ of the order of $\epsilon$ and $\phi_{2}$ of the 
order of $\epsilon^{2}$. Because of the fast decreasing behavior of
$\partial_{i} \phi$ for $\xi > \epsilon$, the terms like 
$\xi \partial_{i} \phi$ are one order of correction higher than 
$\partial_{i} \phi$ and, in the expansion of (14) in powers of $\epsilon$, 
we must also include the expansion of the metric terms in powers of $\xi$.  
{}From (4-7) and (8-11), we can compute $K_{i}$ defined by :

\[ K_{i} = \frac{1}{\sqrt{-g}} \partial_{i} (\sqrt{-g})  = \]

\[ = \frac{1}{2} \left( g^{AB} \partial_{i} g_{AB} +
                      g^{Aj} \partial_{i} g_{Aj} +
					  g^{kB} \partial_{i} g_{kB} +
					  g^{kj} \partial_{i} g_{kj} \right) \]         

We find :

 \begin{equation} 
 K_{i} = K_{i}^{0} +  \xi^{j} K^{1}_{ij} 
 \end{equation}

with

\begin{equation} 
K_{i}^{0} = g^{AB} b_{ABi} 
\end{equation}

\begin{equation}
K_{ij}^{1} = - b_{ABi} b^{AB}_{\,\,\,\,\,\,\,j}
\end{equation}

and from Gauss-Weingarten's equation :

\[ N^{\mu}_{i,A} = g^{CD} b_{ACi} X^{\mu}_{,D} +
                   g^{kj} A_{kiA} N^{\mu}_{j} \]

where $K_{i}^{0}$ is the mean curvature, $K_{ij}^{1}$ is the gaussian
curvature, 				   
$b_{ACi}$ is the second fundamental form and 
$A_{kiA}$ is the twisting vector.
Their values depend on the surface we are dealing with and they must be
compared with the $\partial_{i} \partial^{i} \phi$ - terms.  
They may be large
and important or not, depending on how bent the surface is. Anyway, 
they are not related to the orders of correction of the defect width.
 
Now, using (15) and (12), equation (14) splits into two pieces
to describe the zero and the first order equations. Up to zero
order of $\epsilon$ we have:

\begin{equation}
\partial_{i} \partial^{i} \phi_{0} +
K_{i}^{0} \partial^{i} \phi_{0} +
\left.\frac{\partial V}{\partial \phi}\right\rfloor_{0} = 0
\end{equation}

where $\rfloor_{0}$ indicates evaluation at $\phi = \phi_{0}$. 
The first order correction, in $\epsilon$, of (14) is :

\begin{equation}
\partial_{i} \partial^{i} \phi_{1} +
K^{0}_{i} \partial^{i} \phi_{1} +
K_{ij}^{1} \xi^{i} \partial^{j} \phi_{0} +
\phi_{1} \left.\frac{\partial^{2} V}{\partial \phi^{2}}\right\rfloor_{0} = 0
\end{equation}

{}From (19), we note that $\phi_{1}$ can not be identically zero as long as 
$K^{1}_{ij}$ and $\partial^{j} \phi_{0}$ are not zero.
Up to the order of $\epsilon^{2}$, we would get an equation for $\phi_{2}$.
However, this contribution will not be important for our calculations.

In the next section we will compute the effective action, integrating
out $\xi$ in the original action (13). To do so, it is important that both
$\phi_{0}$ and $\phi_{1}$ are $\sigma^{A}$ - independent. In the standard
derivation of Nambu action, the defect is locally approximated by the static
and straight solution. The world surface has, locally, constant normal vectors
making $b_{ACi} = 0$. Consequently, both $K_{i}^{0}$ and $K^{1}_{ij}$ are
$\sigma^{A}$ - independent (equal to zero) and, from (18-19), $\phi_{0}$
and $\phi_{1}$ are also $\sigma^{A}$ - independent.

In this paper, we consider the defect locally approximated by a non-plane
solution, adjusting the values of $K_{i}^{0}$ and $K_{ij}^{1}$.  
However, to maintain the $\sigma^{A}$- independence of $\phi_{0}$
and $\phi_{1}$, we restrict ourselves to solutions 
with locally constant values of $K_{i}^{0}$
and $K_{ij}^{1}$ . We are approximating
the defect world surface by a local solution surface with constant
mean and gaussian curvatures. This would be satisfied, for example, by
a spherical or cylindrical surface with constant radius. These surfaces
have $b_{ABi} \neq 0 $ and 
$ X^{\mu}_{,A} N^{\nu}_{i,B} G_{\mu\nu} = b_{ABi}$, $\sigma^{A}$- 
independent. In this case, both $\phi_{0}$ and $\phi_{1}$ remain, through
(18-19), $\sigma^{A}$ - independent.

Finally, we should note that the expansion in powers of $\epsilon$ would
not reproduce the results (18) and (19) if it were done directly in the 
action. The Euler- Lagrange equations introduce the derivatives on 
$\xi^{i}$ that change the power counting. We claim that, as long as
we are dealing with the classical dynamical evolution of a function of 
$\xi^{i}$, namely $\phi (\xi^{i})$,  
the correct place to make the expansion in powers of $\xi^{i}$ is the
equation of motion .

\section{The effective action}

The next step is to look for an effective action that will describe
the evolution of the surface $\xi^{i} = 0$ as a geometrical object.
Starting from (13):

\begin{equation}
S = \int \sqrt{-g} {\cal L} (\phi) \, d^{p+1}\sigma \, d^{m} \xi
\end{equation}

the idea is to integrate out the $\xi^{i}$ - dependence, leaving an 
action of the form $S = \int f(\tilde{g}(\sigma)) d\sigma$.
This means that the Euler-Lagrange equations for this effective
action involves neither $\xi^{i}$ nor its derivatives and, 
in order to get the equation of  the defect from the effective action, 
we may safely expand in powers of $\xi^{i}$ directly in the action (20).

We expand ${\cal L} (\phi) = {\cal L}_{0} +
                             {\cal L}_{1} +
							 {\cal L}_{2} $
and replace in (20) to find: 

\[S = \int \sqrt{-g} \left[ 
\frac{1}{2} \partial_{i} \phi_{0} \partial^{i} \phi_{0}
                            - V (\phi_{0}) \right] \,\,
							d^{p+1} \sigma \,\, d^{m} \xi  + \]

\[ + \int \sqrt{-g} \left[
\partial_{i} \phi_{1} \partial^{i} \phi_{0} + 
\partial_{i} \phi_{2} \partial^{i} \phi_{0} - 
\left.\frac{\partial V}{\partial \phi} \right\rfloor_{0} \phi_{1} -
\left.\frac{\partial V}{\partial \phi} \right\rfloor_{0} \phi_{2} \right] 
\,\,  d^{p+1}\sigma \,\, d^{m} \xi + \]
 					
\begin{equation}							
  +  \frac{1}{2} \int \sqrt{-g} \left[
                    \partial_{i} \phi_{1} \partial^{i} \phi_{1} -
		 \left.\frac{\partial^{2} V}{\partial \phi^{2}}\right\rfloor_{0}
		\phi^{2}_{1} \right]
					\,\,	d^{p+1}\sigma \,\, d^{m} \xi
\end{equation}							

Integrating by parts the first and second order correction terms 
 and using equations (18),(19) , we get :

\begin{equation}
S= \int \sqrt{-g} \left[ \left(
          \frac{1}{2} \partial_{i} \phi_{0} \partial^{i} \phi_{0} -
V(\phi_{0})
				         \right) -
	     \frac{1}{2}\xi^{i} K^{1}_{ij} \partial^{j} \phi_{0} \phi_{1}		 
		          \right] \,\,
				  d^{p+1}\sigma\,\,d^{m} \xi
\end{equation}

To be consistent, we must also expand $\sqrt{-g}$ :

\begin{equation}
\sqrt{-g} = \sqrt{-\tilde{g}} +  \partial_{i}(\sqrt{-g})\rfloor_{0}
 \xi^{i} +
\frac{1}{2}  \partial_{i}\partial_{j} (\sqrt{-g})\rfloor_{0}
 \xi^{i} \xi^{j} 
\end{equation}

{}From (4-7),(8-11) and (16-17),
 we find :
 
\begin{equation}  
\partial_{i}(\sqrt{-g})  = \sqrt{-\tilde{g}} K^{0}_{i}
\end{equation}

\[
\partial_{i} \left[ \partial_{j} (\sqrt{-g}) \right] =
\partial_{i} \left[ \sqrt{-g} ( K^{0}_{j} +  \xi^{k} K^{1}_{kj})\right]
 = \]

\begin{equation}
\ =\sqrt{-\tilde{g}} ( K^{0}_{i} K^{0}_{j} + K^{1}_{ij})
\end{equation}

Up to zero order, we have the action:

\begin{equation}
S= \int \sqrt{-\tilde{g}} \left[
\frac{1}{2} \partial_{i}\phi_{0} \partial^{i} \phi_{0} - V(\phi_{0}) 
\right] \,\,
d^{p+1} \sigma \,\, d^{m} \xi
\end{equation}

With $\phi_{0}$ independent of $\sigma^{A}$  
and $\sqrt{-\tilde{g}}$ independent of $\xi^{i}$, the 
integration in $\xi^{i}$ can be performed to reproduce the Nambu action:

\begin{equation}
S = \mu_{0} \int \sqrt{-\tilde{g}} \,\, d^{p+1} \sigma
\end{equation}

and $\mu_{0} = \int d^{m} \xi (
\frac{1}{2} \partial_{i} \phi_{0} \partial^{i} \phi_{0} -
                               V(\phi_{0}))$. As it is well known, this 
describes a minimal surface, with equation	of motion given by :

\begin{equation}
K^{0}_{i} = \tilde{g}^{AB} b_{ABi} = 0
\end{equation}

The first order correction does not vanish on integration over $\xi^{i}$
because $\phi_{0}$, solution of(18), is not an even function of $\xi^{i}$ 
for $K_{i}^{0} \neq 0 $. 
The second order  has two
contributions. The last term in the right hand side of (22) is not zero,
due to the $\phi_{1}$ contribution .
Rewritting these terms , 
and adding the contribution from (23), we have :

\begin{equation}
S = \mu_{0} \int \sqrt{-\tilde{g}} \left[ 1 +
\frac{\mu_{1}^{i}}{\mu_{0}} K_{i}^{0} +
\frac{\mu_{2}^{ij}}{\mu_{0}}  (K_{i}^{0}K_{j}^{0} + K^{1}_{ij}) -
\frac{\tilde{\mu}_{2}^{ij}}{\mu_{0}}  K^{1}_{ij} \right]
 \,\, d^{p+1} \sigma
\end{equation}

with :

\[ \mu_{1}^{i} = \int d^{m} \xi \xi^{i}
\left[ \frac{1}{2} \partial_{j} \phi_{0} \partial^{j} \phi_{0} - 
V(\phi_{0}) \right] \]

\[ \mu_{2}^{ij} = 
\frac{1}{2} \int d^{m} \xi \xi^{i} \xi^{j} \left[ \frac{1}{2}
     \partial_{l} \phi_{0} \partial^{l} \phi_{0} - V(\phi_{0}) \right]  \]

\[ \tilde{\mu}_{2}^{ij} = 
\frac{1}{2} \int d^{m} \xi \xi^{i} \partial^{j}\phi_{0} \phi_{1} \]

Both $\mu_{2}^{ij}$ and $\tilde{\mu}_{2}^{ij}$ are second order contributions
due to $\xi^{2}$ and $\xi \phi_{1}$.

In the specific example  we are dealing with, there is only one direction
$\xi^{i}$ and i takes just one value. We write 
$\tilde{\mu}_{2}^{ij} K_{ij}^{1} = - \tilde{\mu}_{2} K$ with 
$K = b_{ABi} b_{\,\,\,\,\,\,\,\,i}^{AB} $.
Using the Gauss-Codazzi relations :

\begin{equation}
\sum_{i} ((K^{0}_{i})^{2} + K^{1}_{ii}) = - R 
\end{equation}

we get

\begin{equation}
S= \mu_{0} \int \sqrt{-\tilde{g}} \left[ 1 + 
\frac{\mu_{1}}{\mu_{0}} K^{0} -
\frac{\mu_{2}}{\mu_{0}} R +
\frac{\tilde{\mu}_{2} }{\mu_{0}} K \right]
 \,\, d^{p+1} \sigma .
\end{equation}

\section{Conclusion}
So, starting with an arbritary configuration, we calculate the first non-zero
corrections to the generalized Nambu action for a bubble. One term
of this correction agrees with Gregory's result. The others come in only
if the original defect is not plane. They depend on the odd part of 
$\phi_{0}$ and on the field correction
$\phi_{1}$. It should be note that $\phi_{1}$ remains non zero even if we
use the zero order condition $K_{i}^{0}=0$ (28) in the equation of motion of 
$\phi_{1}$, (19). This procedure is advocated by R.Gregory [4] on the basis
of consistency, and for plane zero order solutions it implies identically
zero solutions for $\phi_{1}$. In fact, if the zero order solution is plane,
not only $K_{i}^{0} = 0$ but also $K_{ij}^{1} = 0$ because the normal vectors 
are, up to zero order,constants and satisfy $N_{i,A}^{\mu} = 0$.
 In this case, equation (19) has solutions with $\phi_{1} = 0$.

We should also point out that the choice of pertubations in the form
$\phi = \phi_{0}(\xi^{i}) + \phi_{1} (\xi^{i})$ 
was crucial to allow the factorization of the $\xi$- dependence from
the action integral.
If we had $\phi = \phi( \xi^{i}, \sigma ^{A})$ , it would be impossible
to extract the geometrical part of the dynamics, 
expressed by an action integrated over
$\sigma^{A}$ only, as (31). This restriction means that, if the zero order
is a spherical bubble, the first order correction will not violate
the spherical symmetry.
 
The procedure described here can be applied to other defects.In particular,
for strings, the R term is a topological constant, and the non-trivial 
 correction are K-like terms which 
we expect to be important near cusps.
We should also mention that the existence of a similar term and its influence 
on the rigidity of the string were considered before by Polyakov [6].
The evolution of strings with rigidity produced by extrisinc curvature
terms was  also studied  in [7].In these references, 
an extra contribution to the action proportional to $(K_{i}^{0})^{2}$
is proposed.  
The major difference from our work is that 
instead of putting by hand the  extra term, we obtain this contribution 
directly from the field equations of the theory and, besides,  we get
an additional first order contribution, proportional to the mean curvature
$K^{0}$ and connected with the odd part of the zero order solution $\phi_{0}$.

\section{Acknowledgments}
The authors would like to thank the  CNPq for the finantial support.

\end{document}